\documentclass[aps,pre,twocolumn,floatfix,superscriptaddress,showpacs,amsfonts,amssymb,amsmath,preprintnumbers]{revtex4}
\usepackage{bm}
\usepackage{psfig}
\usepackage{dcolumn}

\newcounter{punkt}
\newcommand{\secref}[1]{Sec.~\ref{#1}}

\newcommand{\Eq}[1]{Equation~(\ref{#1})}
\newcommand{\eq}[1]{Eq.~(\ref{#1})}

\newcommand{\eqsand}[2]{Eqs.~(\ref{#1}) and~(\ref{#2})}

\newcommand{\exref}[1]{(\ref{#1})}
\newcommand{\exsand}[2]{(\ref{#1}) and~(\ref{#2})}

\newcommand{\figref}[1]{Fig.~\ref{#1}}

\newcommand{\bea}{\begin{eqnarray}}
\newcommand{\eea}{\end{eqnarray}}
\newcommand{\bal}{\begin{aligned}}
\newcommand{\eal}{\end{aligned}}
\newcommand{\bga}{\begin{gathered}}
\newcommand{\ega}{\end{gathered}}

\newcommand{\lt}{\left}
\newcommand{\rt}{\right}
\newcommand{\bl}{\bigl}
\newcommand{\br}{\bigr}
\newcommand{\la}{\langle}
\newcommand{\ra}{\rangle}

\newcommand{\const}{\text{const}}
\renewcommand{\phi}{\varphi}

\newcommand{\dd}{\partial}
\newcommand{\diff}{d}

\newcommand{\bigO}{O}

\newcommand{\vu}{{\bf u}}
\newcommand{\vx}{{\bf x}}
\newcommand{\vy}{{\bf y}}
\newcommand{\vk}{{\bf k}}

\newcommand{\kkappa}{{\hat\kappa}}

\newcommand{\ko}{k_\text{flow}}
\newcommand{\kbox}{k_\text{box}}
\newcommand{\kpeak}{k_\text{peak}}
\newcommand{\keta}{k_\eta}
\newcommand{\gL}{\gamma_0}

\newcommand{\CC}{{C_\lambda}}

\newcommand{\Pe}{\text{Pe}} 

\begin{document}

\preprint{E-print {\tt nlin.CD/0404016}; {\em PRE} {\bf 70}, 046304 (2004)}

\title{Diffusion of passive scalar in a finite-scale random flow}

\author{Alexander A.\ Schekochihin}
\email{as629@damtp.cam.ac.uk}
\author{Peter H.\ Haynes} 
\affiliation{DAMTP/CMS, 
University of Cambridge, Wilberforce Road, Cambridge CB3 0WA, UK}
\author{Steven C.\ Cowley}
\affiliation{Department of Physics and Astronomy, 
UCLA, Los Angeles, California 90095-1547, USA}
\affiliation{Plasma Physics Group, Imperial College, 
Blackett Laboratory, Prince Consort Road, London~SW7~2BW, UK}
\date{\today}

\begin{abstract}
We consider a solvable model of the decay of scalar variance 
in a single-scale random velocity field. We show that 
if there is a separation between the flow scale $\ko^{-1}$ 
and the box size $\kbox^{-1}$, 
the decay rate $\lambda\propto(\kbox/\ko)^2$ is determined 
by the turbulent diffusion of the box-scale mode. 
Exponential decay at the rate $\lambda$ is preceded 
by a transient powerlike decay 
(the total scalar variance $\sim t^{-5/2}$ if 
the Corrsin invariant is zero, $t^{-3/2}$ otherwise) 
that lasts a time $t\sim1/\lambda$. 
Spectra are sharply peaked at $k=\kbox$.
The box-scale peak acts as a slowly decaying source 
to a secondary peak at the flow scale. 
The variance spectrum at scales intermediate 
between the two peaks ($\kbox\ll k\ll\ko$) is 
$\sim k + a k^2 +\dots$ ($a>0$). 
The mixing of the flow-scale modes by the random flow 
produces, for the case of large P\'eclet number, 
a $k^{-1+\delta}$ spectrum at $k\gg\ko$, where 
$\delta\propto\lambda$ is a small correction. 
Our solution thus elucidates the spectral make up 
of the ``strange mode,'' combining small-scale 
structure and a decay law set by the largest scales. 
\end{abstract}


\maketitle

\section{Introduction}

The problem of the decay of passive-scalar variance has 
recently been reexamined in the literature 
following the realization that the decay rates, spectra, and 
higher-order statistics based on small-scale 
Lagrangian-stretching theories 
\cite{Antonsen_etal,Son,Balkovsky_Fouxon,Falkovich_etal_review} 
are not consistent with either numerical 
\cite{Pierrehumbert_strange_mode,Pierrehumbert_pdfs,Fereday_etal,Sukhatme_Pierrehumbert} 
or experimental \cite{Rothstein_etal,Voth_etal} results 
in the long-time limit. 
Instead, the scalar decay is dominated 
by an eigenmodelike solution dubbed ``the strange mode'' 
\cite{Pierrehumbert_strange_mode}
because it combines intricate small-scale structure 
with globally determined decay rate and self-similar 
statistics (self-similarity is also seen in numerical 
simulations of the related problem of kinematic dynamo \cite{SCMM_ssd}). 
There has been a growing understanding 
\cite{Voth_etal,Thiffeault_Childress,Chertkov_Lebedev,Fereday_Haynes,Sukhatme_pdf} 
that the overall decay rate is set by the slowest-decaying 
system-scale modes. This brings to mind homogenization 
theory \cite{Majda_Kramer}, 
which considers the turbulent diffusion of passive scalar 
at scales much larger than the flow scale and where 
it is the largest-scale mode that decays most slowly. 
In this paper, we use a simple solvable example 
to demonstrate that the strange-mode decay rate is 
the rate of turbulent diffusion of the box-scale 
mode and show how the spectra of scalar variance accommodate 
both this box-scale diffusion and small-scale structure. 

Qualitatively, the key idea quantified by our theory 
is as follows. A scalar field whose variance is at the scale 
smaller than or equal to the scale of the ambient random flow 
is mixed at a rate determined by the Lyapunov exponent
of the flow --- this is the Lagrangian-stretching approach. 
However, if the size of the box is larger than the 
scale of the flow, the scalar field can have variance 
at the scale of the box. The rate of transfer of this 
large-scale variance to the flow scale (turbulent diffusion) 
can be much smaller than the Lagrangian mixing rate, 
in which case this slow transfer sets the global decay rate. 

Our model emphasizes scale separation between the box 
and the flow. 
Our results are complementary to \cite{Haynes_Vanneste}, 
where the decay of a scalar field is studied 
with more generality (in two dimensions).

We consider the advection-diffusion equation 
\bea
\label{ADEq}
\dd_t\theta + \vu\cdot\nabla\theta = \eta\Delta\theta,
\eea
with a random Gaussian white-in-time velocity field 
$\la u^i(t,\vx)u^j(t',\vx')\ra = \delta(t-t')\kappa^{ij}(\vx-\vx')$
known as the Kraichnan model \cite{Kraichnan1}. 
The mean scalar concentration has been 
subtracted --- i.e., $\la\theta\ra=0$. 
For the Kraichnan velocity, 
the angle-integrated scalar-variance spectrum 
in $d$ dimensions 
$T(k)=\int\diff\Omega_\vk\, k^{d-1} \la|\theta(\vk)|^2\ra$
satisfies an integro-differential equation valid 
at all~$k$~\footnote{The 
derivation is analogous to the standard 
one in the dynamo theory: see, e.g., \cite{SBK_review} 
and references therein. Note that the $x$ space version 
of \eq{MCEq} is local, but we stay with the integral 
equation because we are interested in spectra. 
For $x$ space calculations, see 
\cite{Chertkov_Lebedev,Haynes_Vanneste}.}: 
\bea
\nonumber
\dd_t T(t,k) &+& (2\eta + \kappa_0) k^2 T(t,k) \\ 
\label{MCEq}
&=& k_i k_j\int\diff^d k'\kappa^{ij}(\vk-\vk')\,T(t,k'),
\eea
where $\kkappa^{ij}(\vk)=\kkappa(k)\bl(\delta^{ij}-k_i k_j/k^2\br)$ 
is the Fourier transform of $\kappa^{ij}(\vx-\vx')$ 
and $\kappa_0=(1/d)\kappa^{ii}(0)$ is the turbulent diffusivity 
($d$ is the dimension of space). 

In \secref{sec_small_scales}, we review the theory of scalar 
decay at small scales, which leads to the standard Lagrangian-stretching 
results. In \secref{sec_main}, the theory for a finite-scale 
flow is developed --- this is the main part of the paper. 
Concluding remarks are in \secref{sec_disc}. 

\section{Small-scale theory} 
\label{sec_small_scales}

If the P\'eclet number $\Pe=u\ko^{-1}/\eta$ is large, $\theta$ 
varies at scales as small as $\Pe^{-1/2}$ times the scale of the flow. 
This small-scale structure can be considered in the 
approximation of spatially 
linear velocity field \cite{Batchelor,Kraichnan2}, 
---~viz., 
\bea
\label{kappa_exp}
\kappa^{ij}(\vy) \simeq \kappa_0\delta^{ij} 
- {1\over2}\,\kappa_2\lt(y^2\delta^{ij} - {1\over2}\,y^i y^j\rt).
\eea
For the Kraichnan velocity, the Lagrangian-stretching theories 
\cite{Antonsen_etal,Son,Balkovsky_Fouxon,Falkovich_etal_review} 
amount to the approximation~\exref{kappa_exp}. 
The scalar-variance spectrum 
satisfies a Fokker-Planck-type equation 
\cite{Kraichnan1,Kraichnan2} 
\bea
\label{FPEq}
\dd_t T = D\,{\dd\over\dd k}
\lt[k^2{\dd T\over\dd k} - (d-1)k T\rt] -2\eta k^2 T,
\eea
where $D=[(d-1)/2(d+1)]\kappa_2$. 
This equation is valid for $k\gg\ko$. 
In this limit, it either can be obtained from \eq{MCEq} 
by expanding the mode-coupling term on the right-hand side 
or derived directly by assuming linear velocity field \cite{Kraichnan2}. 

The solution of \eq{FPEq} that decays at $k\to\infty$ is 
\bea
\label{FPSln}
T(t,k) = \CC\,e^{-\lambda\gL t} k^{-1+d/2} 
K_{(d/2)\sqrt{1-\lambda}}\,(k/\keta),
\eea
where $\CC$ is a constant, 
$K_\nu(z)$ is the modified Bessel function of the second kind, 
$\gL=(d^2/4)D$, $\keta=(D/2\eta)^{1/2}$, 
and $\lambda$ is the nondimensionalized decay rate, 
which must be calculated by applying the correct boundary 
condition at small $k$. 
If we assumed that the decay rate is fully determined 
by the small scales, a reasonable procedure would be to 
choose some infrared cutoff $k_*$ and require 
the flux of scalar variance through $k_*$ [the 
square brackets in \eq{FPEq}] to vanish 
(cf.~\cite{SCHMM_ssim}). This can 
be satisfied only for $\lambda>1$. For $k_*\ll\keta$, 
the zero-flux condition becomes 
$\sin\bl[(d/2)\sqrt{\lambda-1}\,\ln(k_*/2\keta)\br] = 0$.
Placing the cutoff $k_*$ at the largest zero, 
ensures that $T(k)$ is everywhere positive. 
We get (cf. \cite{Haynes_Vanneste})
\bea
\label{log2_formula}
\lambda = 1 + {(2\pi/d)^2\over\bl[\ln(k_*/2\keta)\br]^2} 
= 1 + \bigO\bl(1/\log^2\Pe\br).
\eea
This implies a $k^{(d-2)/2}$ spectrum at $k\gg\keta$. 
If the scalar variance is initially at $k\gg\ko$, 
these results (or their analogs for other model flows) 
hold during the initial stage of the scalar decay. 
However, in the long-time limit 
for cases in which the system (box) size is 
(several times) larger than the flow scale, 
both numerical simulations \cite{Fereday_Haynes}
and experimental results \cite{Voth_etal} 
obtain much smaller decay rates and spectra with negative exponents. 
The conclusion is that the zero-flux boundary condition 
is incorrect and the decay rate $\lambda<1$ 
must be determined by matching 
the solution \exref{FPSln} to the solution at nonlarge $k$ 
where \eq{FPEq} is invalid and \eq{MCEq} must be used instead. 
The spectrum at $\ko\ll k\ll\keta$ 
is then $\sim k^{s(\lambda)}$, where 
[from \eq{FPSln}]
\bea
\label{s_formula}
s(\lambda) = -1 + (d/2)\bl(1-\sqrt{1-\lambda}\br).
\eea
Note that for $\lambda\ll1$, 
$s(\lambda)\simeq -1 + (d/4)\lambda$, which coincides 
with the formula proposed in \cite{Fereday_Haynes}. 

For the initial spectrum concentrated at $k\sim\keta$, 
the period of validity of \eq{log2_formula} is the time 
it takes the spectrum to spread to $k\sim\ko$. 
The spreading can be shown to be exponentially fast 
with a rate $\sim\gL$. Since $\keta\sim\Pe^{1/2}\ko$, 
the Lagrangian-stretching results are 
valid for $t\ll\gL^{-1}\log\Pe$. 

\section{Theory for a finite-scale flow} 
\label{sec_main}

The challenge now is to find $\lambda$ by 
solving \eq{MCEq}. 
Let us specialize to three dimensions 
(theory in 2D is analogous) 
and choose a simple form for the velocity 
correlator: $\kkappa(k)=N\,\delta(k-\ko)$, 
where $N = 15\kappa_2/16\pi\ko^4$ 
[note that $\kappa_2=(2/5)\ko^2\kappa_0$]. 
This describes a Kraichnan ensemble 
of randomly oriented ``eddies'' of size $\ko$. 
We set $\ko=1$ and carry out angle integrations in \eq{MCEq} 
to get  
\bea
\nonumber
\dd_t T(t,k) &+& (2\eta + \kappa_0) k^2 T(t,k) \\ 
&=& {15\over32}\,\kappa_2\, k \int_{|1-k|}^{1+k}{\diff k'\over k'}\,
K(k,k') T(t,k'),
\label{Eq3D}
\eea
where $K(k,k')=-k^{\prime4} + 2(1+k^2)k^{\prime2}-(1-k^2)^2$. 
It is not hard to ascertain that \eq{Eq3D} reduces 
to \eq{FPEq} when $k\gg1$. 
Let us now consider the opposite limit $k\ll1$ --- i.e., 
the evolution of scale variance at scales much larger 
than the scale of the flow. In this limit, the integral 
in \eq{Eq3D} is dominated by the modes in the neighborhood
of the flow scale $k=1$. Neglecting $\eta$, we get 
\bea
\label{HomEq}
\dd_t T + k^2 T = {3\over4}\,k\int_{-k}^k
\diff q\,(k^2-q^2)T(t,1+q) \equiv S(t,k),
\eea
where time is rescaled $t\kappa_0\ko^2\to t$. 
The solution is 
\bea
\label{HomSln}
T(t,k) = \lt[T(0,k) + 
\int_0^t\diff t' S(t',k) e^{k^2 t'}\rt] e^{-k^2 t}.
\eea
In order to complete the solution, we must determine 
$T(t,1+q)$. For $q\ll1$, it satisfies 
\bea
\label{T1_eq}
\dd_t T + T = 
{3\over4}\,\int_{|q|}^{2+q}{\diff k'\over k'}
\lt(k^{\prime2}-{k^{\prime4}\over4}-q^2\rt) T(t,k'). 
\eea

As we shall see, the solution~\exref{HomSln} is sharply peaked 
at $k=\kpeak\sim1/\sqrt{t}$. 
In an infinite system, this peak would move 
indefinitely towards ever smaller~$k$. 
In a finite system, $\kpeak$ eventually becomes 
comparable to the inverse system size. 
Strictly speaking, this means that 
one must solve the problem with a discrete set of modes  
and application-specific boundary conditions 
(cf.~\cite{Chertkov_Lebedev,Lebedev_Turitsyn,Haynes_Vanneste}). 
Instead, we model the finite box 
by introducing an infrared cutoff $\kbox$ into our 
continuous theory. All lower integration limits in $k$ space 
are subject to this cutoff. This is not a rigorous 
operation, but it is a reasonable modeling choice 
as long as $\kbox\ll1$. The time at which $\kpeak\sim\kbox$ 
is $t\sim1/\kbox^2$. Therefore, there will be two asymptotic regimes 
of scalar decay: {\em the transient stage} $t\ll1/\kbox^2$, 
when unmodified continuous theory can be used, 
and {\em the long-time limit} $t\gg1/\kbox^2$, 
when the box cutoff is important. 
Note that even for the transient stage, we assume 
$t\gg1$; i.e., we consider times much longer 
than the ``turnover time'' of the flow. 
For spectra initially at small scales, we harden 
this condition to $t\gg\log\Pe$, so that 
the Lagrangian-stretching theory ceases to be valid. 

Consider \eq{HomSln}. 
Since $k\ll1$, we can Taylor-expand the initial spectrum: 
$T(0,k)=C_2 k^2 + C_4 k^4 +\dots$. 
Here $C_2$ is the Corrsin invariant, 
$C_2\propto\int\diff^3 x\,\la\theta(\vx)\theta(0)\ra$ \cite{Corrsin_inv}. 
Some aspects of the scalar decay differ for cases 
with $C_2>0$ and $C_2=0$. We shall first develop our 
theory for $C_2>0$. The case of $C_2=0$
will be treated in \secref{sec_C2_zero}. 

\subsection{Case of $C_2>0$}
\label{sec_C2}

Consider first the transient stage $1\ll t\ll1/\kbox^2$. 
Let us assume that the dominant term in the 
solution~\exref{HomSln} is 
\bea
\label{HomSln_C2}
T(t,k) = C_2 k^2 e^{-k^2 t}, 
\eea
which peaks at $\kpeak=1/\sqrt{t}$. 
We shall justify this assumption {\em a posteriori}. 
Let us now determine the flow-scale solution [\eq{T1_eq}]. 
Assuming that the main contribution to the 
integral in \eq{T1_eq} is from $k'\ll1$ 
(also to be verified later) and substituting \eq{HomSln_C2} 
for $T(t,k')$, 
we get, to leading order in in $1/t$ (and neglecting $\dd_t T$), 
\bea
\label{T1_decay_C2} 
T(t,1+q) = {3\over8}\,{C_2\over t^2}\,e^{-q^2 t} \equiv T_1(t) e^{-q^2 t},
\eea
where $T_1(t)=T(t,1)$.
This describes the neighborhood of the flow scale, where 
the coupling to the small-$k$ modes produces a secondary peak 
with the width $\sim1/\sqrt{t}$. We shall see below 
that \eq{T1_decay_C2} is, in fact, valid 
beyond the width of the peak and up to $|q|\sim(\ln t/t)^{1/2}$. 
We now substitute \eq{T1_decay_C2} back into \eq{HomSln} 
to see that \eq{HomSln_C2} is, indeed, the dominant solution. 

For $k\ll1/\sqrt{t}$, 
we get $S(t,k)\simeq k^4 T_1(t)$. 
\Eq{HomSln} becomes, to two leading orders in $k$, 
\bea
\label{HomSln_small_k}
T(t,k) \simeq \lt[C_2 k^2 + 
\lt(C_4 + \int_0^t\diff t' T_1(t')\rt)k^4\rt] e^{-k^2 t}. 
\eea
Since $T_1(t)$ decays faster than $1/t$ [\eq{T1_decay_C2}], 
its time integral in \eq{HomSln_small_k} tends to a constant 
when $t\gg1$ and is dominated by the initial stage of 
the evolution of $T_1(t')$ [there is no divergence at $t'=0$ 
because the solution~\exref{T1_decay_C2} is only valid for $t'\gg1$]. 
This time integral represents a finite amount of scalar 
variance that is initially transferred from the flow scale 
to the large scales (small $k$). 
That the effect of nonlinear coupling only appears 
in the $k^4$ term is a reflection 
of the conservation of the Corrsin invariant: 
the coefficient in front of $k^2$ cannot be changed. 
We see that, as long as $C_2\gg\const\times\kbox^2$, 
the first term in \eq{HomSln_small_k} dominates. 
Note that for $k\sim1/\sqrt{t}$, it is still true that 
the time integral in \eq{HomSln} is $\sim k^4$, so 
the above estimates remain valid. 

Now consider $1/\sqrt{t}\ll k\ll1$. 
In this limit, $S(t,k)\simeq(3/4)\sqrt{\pi}\,T_1(t)t^{-1/2}k^3$.
Substituting into \eq{HomSln}, we get 
\bea
\label{Tint_decay_C2}
T(t,k)\simeq C_2 k^2 e^{-k^2 t} 
+ {3\over4}\,\sqrt{\pi}\, T_1(t)t^{-1/2}k. 
\eea
The second term becomes comparable to the first 
at $k\sim(\ln t/t)^{1/2}$. At larger $k$ (but still $k\ll1$), 
it replaces the solution~\exref{HomSln_C2} as the dominant asymptotic. 
The solution~\exref{Tint_decay_C2} is uniformly valid 
across the transition region. Together 
with \eq{T1_eq}, this implies that the solution~\exref{T1_decay_C2} 
is valid for $|q|\lesssim (\ln t/t)^{1/2}$. 

In the long-time limit $t\gg1/\kbox^2$, 
\eq{HomSln} is still valid. 
Again, we assume that the dominant solution is \eq{HomSln_C2}. 
Its peak is at $\kpeak=\kbox$. 
In \eq{T1_eq}, the lower integration limit is adjusted to 
$\max\{|q|,\,\kbox\}$, as explained above. 
The flow-scale peak is now confined 
to $|q|<\kbox$. At these $q$, \eq{T1_decay_C2} is replaced by  
\bea
\label{T1_mode_C2}
T(t,1+q) = {3\over8}\,{C_2\over t}\,(\kbox^2-q^2)\,e^{-\kbox^2 t}.
\eea
This solution gives the dominant contribution to $S(t,x)$
[\eq{HomEq}], so we have (integrating from $-\kbox$ to $\kbox$) 
$S(t,k)=(3/8)C_2 t^{-1} k(k^2-\kbox^2/5)\kbox^3\exp(-\kbox^2 t)$, 
which we substitute into \eq{HomSln}.
The box mode obeys 
\bea
\label{Tbox_mode_C2}
T(t,\kbox) \simeq \lt(C_2\kbox^2 + {3\over10}\,C_2\kbox^6\ln t\rt)
e^{-\kbox^2 t}.
\eea
The time integral now has a logarithmic divergence, representing 
a small amount of transfer of scalar variance from 
the flow-scale mode to the box mode. This contribution 
is not significant because  
the second term in \eq{Tbox_mode_C2} only exceeds the 
first at $t=\exp(10/3\kbox^4)$, which is unphysically large even 
for moderately small values of $\kbox$. 
The width of the box-scale peak, for which the 
decay law~\exref{Tbox_mode_C2} is valid, is estimated 
by $k^2-\kbox^2\lesssim 1/t$ --- i.e., $k-\kbox\lesssim1/2\kbox t$. 

Outside the peak $(k^2-\kbox^2)t\gg1$, we have 
\bea
\nonumber
T(t,k) &\simeq& C_2 k^2 e^{-k^2 t}\\ 
&& +\,\, {3\over 8}\,{C_2\over t}\,
k\,{k^2-\kbox^2/5\over k^2-\kbox^2}\,\kbox^3 e^{-\kbox^2 t}. 
\eea
The first and second terms are of the same order when 
$k^2-\kbox^2\sim\ln t/t$ --- i.e., $k-\kbox\sim\ln t/\kbox t$. 
In the intermediate scale range $\kbox\ll k\ll1$, we have 
\bea
\label{Tint_mode_C2}
T(t,k) \simeq {3\over 8}\,{C_2\over t}\,k \kbox^3 e^{-\kbox^2 t} 
= T_1(t)\kbox k. 
\eea

\begin{figure*}[t]
\centerline{\psfig{file=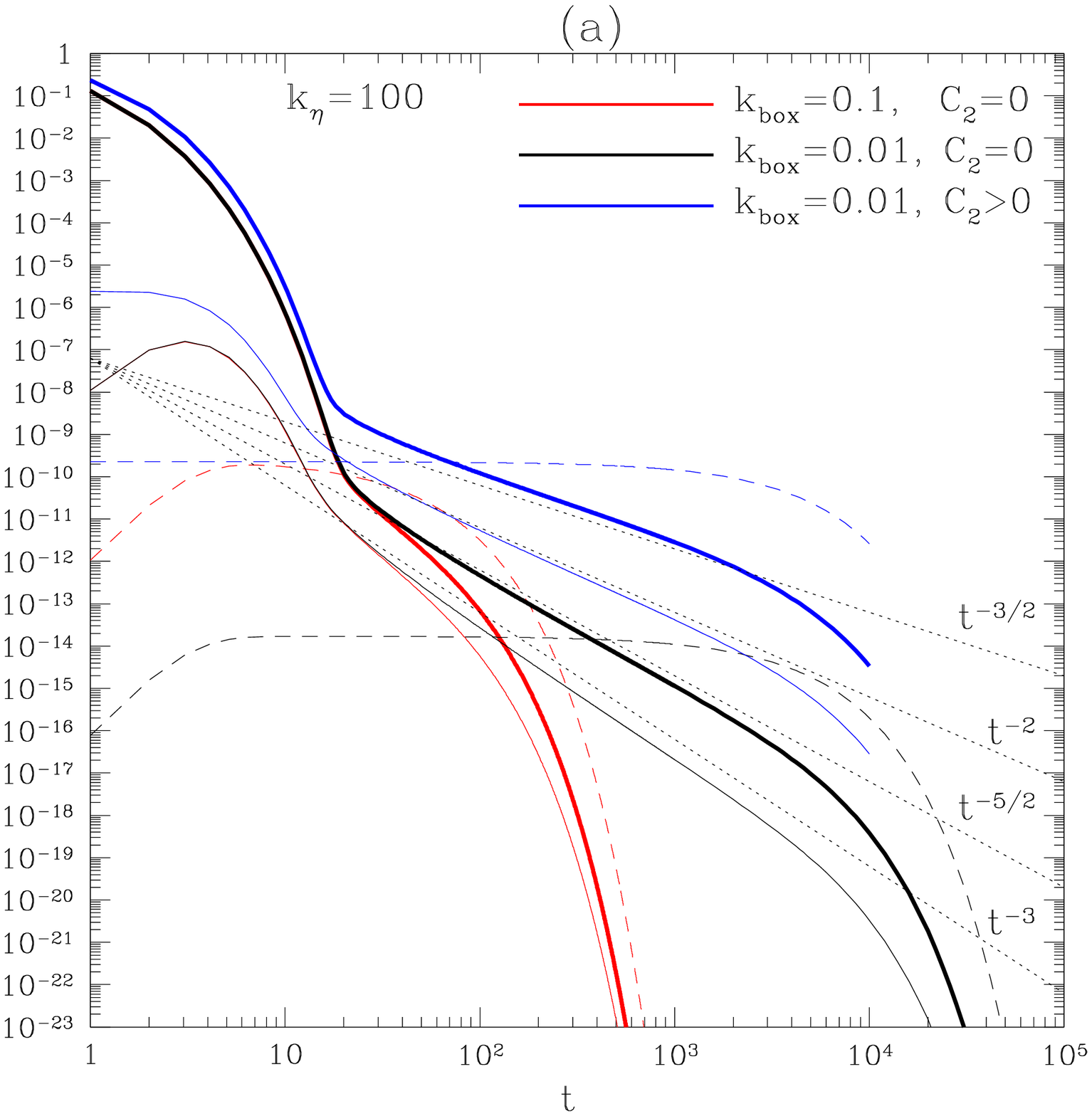,width=9cm}
\psfig{file=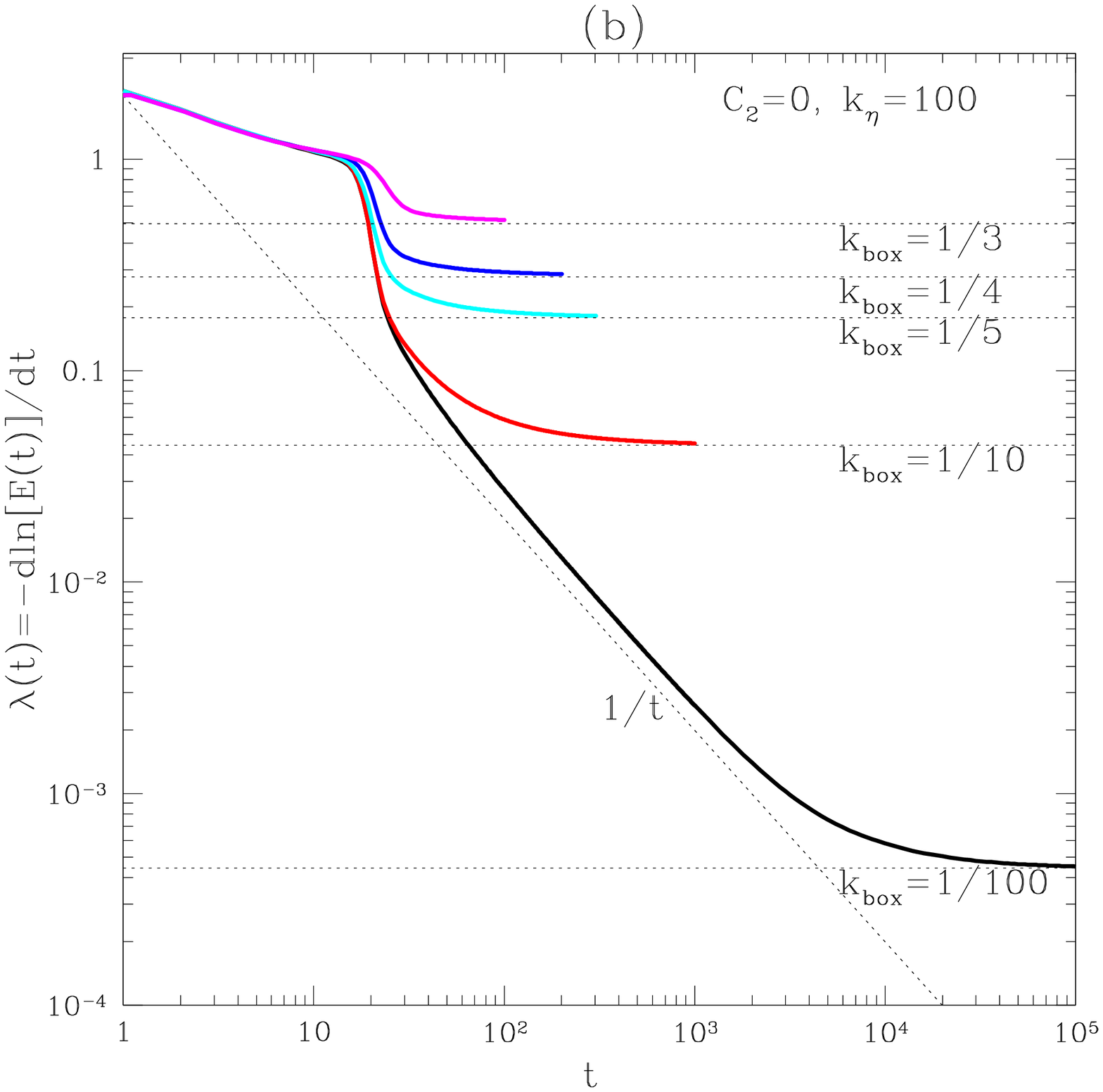,width=9cm}} 
\caption{\label{fig_Et} (a) Scalar-variance decay: 
bold lines depict the total variance $E(t)$, 
thin lines the flow-scale mode $T(t,1)$, 
and dashed lines the box mode $T(t,\kbox)$. 
Dotted lines are theoretical slopes. 
Time is in units of $\gL^{-1}$. 
The $C_2=0$ runs are the same as in \figref{fig_Tk}. 
(b) Effective decay rate $\lambda(t)=-\diff\ln E(t)/\diff t$. 
in units of the Lagrangian-stretching decay rate~$\gL$ 
for various values of $\kbox$. 
Dotted lines correspond to $\lambda=(40/9)\kbox^2$.
The initial spectrum was the same as for \figref{fig_Tk}.} 
\end{figure*}

Let us summarize what we have learned so far. 
We have been concerned with two narrow bands of modes: 
the flow-scale modes $T(t,1+q)$, $q\ll1$, and 
the large-scale peak $T(t,\kpeak)$ at $\kpeak=1/\sqrt{t}$, 
which became the box mode $T(t,\kbox)$ in the long-time 
limit $t\gg1/\kbox^2$. 
The width of the peak was $\sim1/\sqrt{t}$ when $t\ll1/\kbox^2$ 
and $\sim 1/\kbox t$ when $t\gg1/\kbox^2$. 
The flow-scale modes could be assumed to be coupled 
solely to the peak because of the peak's sharp dominance  
of all other modes: indeed, $T(t,\kpeak)\sim t T_1(t)\gg T_1(t)$. 
The width of the secondary peak at the flow scale 
was determined by $\kpeak$. This width can be parametrized 
by $\Lambda(t)=(\kbox^2 + 1/t)^{1/2}$. 

The large-scale peak and the flow-scale band 
are singularities of the scalar-variance spectrum. 
Note that in the long-time limit $t\gg1/\kbox^2$, 
the width of the flow-scale peak is $\Lambda=\kbox$, 
while the width of the box mode is $1/\kbox t\ll\kbox$.
In a finite system, the spacing of the modes cannot be 
smaller than $\kbox$, so it is, of course, unphysical to talk 
about variation of the spectrum at distances less than $\kbox$. 
In our continuous theory, the collapse of the singularities 
to profiles narrower than $\kbox$ means that they should 
be interpreted as single modes. 
Formally, they can be represented as $\delta$ functions: 
thus, for the flow-scale band $1-\Lambda(t)<k<1+\Lambda(t)$, 
we write $T(t,k) = N_1 T_1(t)\Lambda(t)\delta(k-1)$, 
where $N_1$ is a constant arising from 
integrating over the specific shape of the singularity: 
$N_1 = \sqrt{\pi}$ in the transient stage and 
$N_1 = 4/3$ in the long-time limit.

\subsection{Nonsingular spectrum}
\label{sec_nonsing}

Let us now consider the nonsingular part of the spectrum. 
For all $k>\Lambda(t)$, we write 
\bea
T(t,k) = N_1\Lambda(t)T_1(t)\bl[\delta(k-1)+f(k)\br].
\eea 
Substituting this decomposition into 
\eq{Eq3D} and neglecting 
$\diff\ln[\Lambda(t)T_1(t)]/\diff t$, we find 
the equation for $f(k)$: 
\bea
\nonumber
\lt(1+{1\over10\,\keta^2}\rt)f(k) &=& {3\over4}\,k\lt(1-{k^2\over4}\rt)H(2-k)\\ 
&+&{3\over16}\,{1\over k}\int_{|1-k|}^{1+k}
{\diff k'\over k'}\,K(k,k')f(k'),\quad\quad
\label{f_eq}
\eea
where $H(2-k)$ is the Heaviside step function 
[the term it multiplies comes from integrating 
$\delta(k-1)$ and vanishes for $k>2$]. 
For $k\ll1$, we expand $f(k')$ under the integral 
around $k'=1$ and find that the lowest-order 
term is $f(1) k^2$. Clearly, $f(1)>0$. 
Thus, the spectrum at scales intermediate between 
the box scale and the flow scale is
\bea
\label{T_series}
T(t,k) = N_1\Lambda(t)T_1(t)
\bl[(3/4)k + f(1)k^2 + \dots\br].
\eea
Coefficients in higher-order terms  will 
involve derivatives of $f$ at $k=1$. 
The first term in the expansion~\exref{T_series} 
comes from coupling to the flow-scale mode and 
is consistent with the asymptotics~\exsand{Tint_decay_C2}{Tint_mode_C2},  
so the nonsingular solution connects smoothly to the 
large-scale peak. 
The interaction of the nonsingular modes between themselves 
enters in the second order. 

To complete the solution, we would have to find 
$f(k)$ at $k\ge1$. For $k\le2$, this means solving 
the full integral equation, but we do not really need 
to do this. Note that 
the modes with $k>2$ are not directly coupled 
to the flow-scale mode. 
In a rough way, it can be said that the Fokker-Planck 
regime starts at $k=2$ and the solution of \eq{FPEq} 
must be matched to $T(t,2)=N_1\Lambda(t)T_1(t)f(2)$.
The specific value of $f(2)$ 
is not important. The matching is done 
by setting $\CC \propto \Lambda(t)T_1(t)$ 
and $\lambda=-\gL^{-1}\,\diff\ln[\Lambda(t)T_1(t)]/\diff t$ 
(the effective decay rate) in \eq{FPSln}. 
Here $\gL=(9/16)\kappa_2=(9/40)\ko^2\kappa_0$ 
is the decay rate from the Lagrangian-stretching theory. 
During the transient stage, $\lambda\sim1/t$. 
In the long-time limit, $\lambda=(40/9)(\kbox/\ko)^2\ll1$. 
The spectral exponent at $k\gg1$ is given
by \eq{s_formula} and is only slightly shallower than $-1$. 
The spectrum at small scales is, thus, very similar 
to the Batchelor spectrum for the forced scalar 
turbulence \cite{Batchelor}. The physical reason for this 
similarity is that the modes at the flow scale and above 
act as a slowly decaying source to the small-scale part 
of the spectrum, thus making the small-scale physics similar 
to the forced case (cf.~\cite{Fereday_Haynes}).

\subsection{Decay of scalar variance ($C_2>0$)}
\label{sec_total}

Finally, we estimate the decay of the total scalar variance. 
The total variance is the integral of the wave-number spectrum, 
which is made up of the nonsingular 
spectrum [\eq{f_eq}] and the two singular peaks. 
To obtain the contribution of the latter to the 
total variance, we must take into account their 
time-dependent widths. We see that the contributions 
from the flow-scale band and from the nonsingular spectrum 
are always of the same order $\sim\Lambda(t)T_1(t)$ 
[taking into account that the spectrum 
at $k\gg1$ is $f(k)\sim k^{-1+(3/4)\lambda}$
up to $\keta\sim\Pe^{1/2}$ gives an extra factor 
of $\Pe^{(3/8)\lambda}$ to the nonsingular contribution]. 
Therefore, during the transient stage ($t\ll1/\kbox^2$), 
this part of the variance decays as $1/t^{5/2}$. 
In the long-time limit ($t\gg1/\kbox^2$), we have 
$\Lambda(t)T_1(t)\sim \kbox T_1(t)\propto t^{-1}\exp(-\lambda t)$. 
The large-scale peak always decays as $t\,T_1(t)$. 
In the transient stage, its width is $\sim 1/\sqrt{t}$, 
which means that its contribution 
$\sim 1/t^{3/2}$ dominates the rest of the spectrum 
by a factor of $\sim t$ and determines the overall decay 
law of the total variance \footnote{Note that the transient-stage 
powerlike decay laws for the scalar variance can, in fact, 
be obtained from purely dimensional considerations: see, e.g., 
\cite{Chasnov} and references therein. The same is true 
about the decay laws for the case of $C_2=0$ derived 
in \secref{sec_C2_zero}.}. 
In the long-time limit, the width of the box mode is 
$\sim1/\kbox t$, so its decay law is $\sim T_1(t)/\kbox$; i.e., 
it has the same time dependence as the rest of the spectrum, 
but its contribution to the total variance exceeds 
that of all other modes by a factor of $\sim1/\kbox^2$. 

Note that these arguments can be tested for consistency 
with the conservation law for the scalar variance in the 
following way. Pick some wave number $1\ll k\ll\keta$. 
Because of the dominance of the box mode, the scalar variance 
integrated up to $k$ is the same as the total variance. 
Its time derivative must be equal to the flux of variance 
through $k$ [the negative of the expression in square brackets 
in \eq{FPEq}] because dissipation is negligible at $k\ll\keta$. 
It is easy to check that this, indeed, holds true. 
This argument emphasizes that the turbulent diffusion 
at the box scale, which we have shown to control scalar decay, 
is not a dissipation mechanism, but rather describes transfer 
of scalar variance to small scales, where all the dissipation 
is done by molecular diffusion.

\begin{figure}[t!]
\centerline{\psfig{file=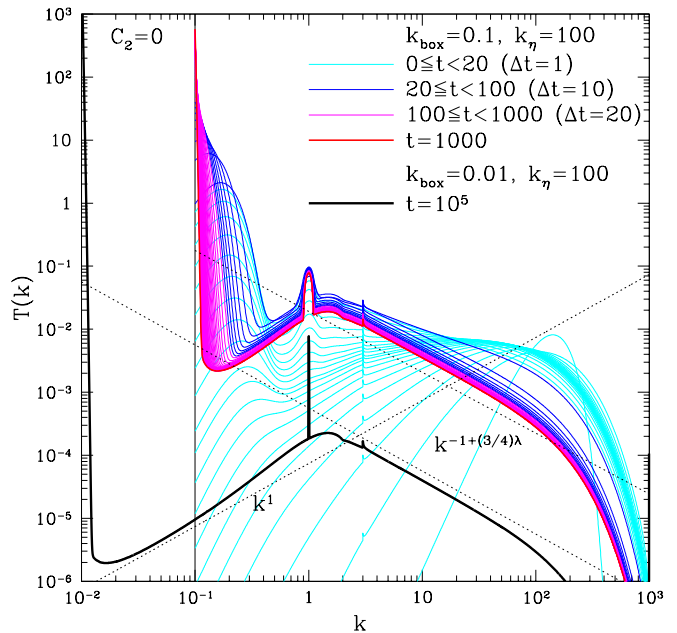,width=9cm}}
\caption{\label{fig_Tk} Evolution of 
normalized spectrum $T(t,k)/E(t)$ for $\kbox=0.1$. 
The initial spectrum was $T(0,k)=k^4\exp[-(k/\keta)^2]$. 
The long-term solution for $\kbox=0.01$ is also shown. 
Spectra at $k<1$ are steeper than $k^1$ 
due to higher-order corrections: 
they can be well fitted by polynomials in the 
form $k + a k^2 + \dots$ [\eq{T_series}]. 
The blip at $k=3$ is the point where the solutions 
of \eq{Eq3D} at $k<3$ and of \eq{FPEq} at $k>3$ 
are spliced together. 
This device allows us to achieve higher $\Pe$. 
We have checked that moving around the splicing 
wave number or solving the full integral equation in the 
entire domain (at lower $\Pe$) do not change the solution.}
\end{figure}

\subsection{Case of $C_2=0$}
\label{sec_C2_zero}

When the Corrsin invariant is zero, the theory developed above 
has to be modified somewhat. The dominant term in the small-$k$ 
solution is $\sim k^4$ and includes contributions both from the 
initial spectrum and from the initial finite amount of 
nonlinear transfer of scalar variance from the flow-scale mode: 
\Eq{HomSln_C2} is replaced by 
\bea
\label{HomSln_zeroC2}
T(t,k) &=& \lt[C_4 + \Phi(t,k)\rt] k^4 e^{-k^2 t},\\
\Phi(t,k) &=& {3\over4}\int_0^t\hskip-1mm\diff t' e^{k^2 t'}
\hskip-1mm\int_{-1}^1\hskip-1mm\diff z (1-z^2) T(t',1+kz).\quad
\eea
In the transient stage ($1\ll t\ll1/\kbox^2$), we have, 
at $k\ll1/\sqrt{t}$, 
$\Phi(t,k)\simeq\int_0^t\diff t' T_1(t')\to\const$ 
as long as $T_1(t)$ decays faster than $1/t$ 
[cf.~\eq{HomSln_small_k}]. Since 
$\Phi(t,k)$ also remains finite for $k\sim 1/\sqrt{t}$, 
we can estimate the flow scale modes $T(t,1+q)$ by 
substituting the solution~\exref{HomSln_zeroC2} into \eq{T1_eq} 
and assuming that $\Phi(t,k)$ is approximately constant 
at the values of $k$ that matter. 
This will produce errors of order unity in the prefactors but 
yield correct scalings and asymptotic decay laws. 
Note that because the nonlinear-transfer 
contribution is $\sim k^4$, the higher-order terms in 
the Taylor expansion for the initial spectrum do not 
affect the asymptotic solutions. 

Further derivation for the $C_2=0$ case follows the same 
general scheme as the theory for $C_2>0$. 
In the transient stage, 
\eq{T1_decay_C2} is replaced by 
\bea
\label{T1_decay_zeroC2}
T(t,1+q) \sim {2 + q^2 t\over t^3}\,e^{-q^2 t},
\eea
where we have dropped prefactors of order unity. 
Note that the transient-stage decay of the flow-scale peak, 
$T_1(t)\sim 1/t^3$, is faster 
then for the $C_2>0$. This is the only important 
change that results from the vanishing of $C_2$. 
The intermediate asymptotic ($1/\sqrt{t}\ll k\ll1$) 
is again $T(t,k)\sim T_1(t) t^{-1/2} k$ [cf.~\eq{Tint_decay_C2}]. 
In the long-time limit, 
\eq{T1_mode_C2} is replaced by 
\bea
T(t,1+q) \sim {\kbox^2\over t}\,(\kbox^2-q^2)\,e^{-\kbox^2 t}, 
\eea
the box mode decays according to [cf.~\eq{Tbox_mode_C2}]
\bea
T(t,\kbox) \sim \kbox^4 e^{-\kbox^2 t}\lt[C_4 + \bigO(\kbox^4\ln t)\rt],
\eea
and the intermediate asymptotic at $\kbox\ll k\ll1$ is 
$T(t,k)\sim T_1(t)\kbox k$ [cf.~\eq{Tint_mode_C2}]. 

The developments for the nonsingular spectrum and for the total 
scalar variance are exactly as described in \secref{sec_nonsing} 
and \secref{sec_total}. The only change is in the transient-stage 
decay laws due to faster decay of the flow-scale 
mode [\eq{T1_decay_zeroC2}]: the contribution to the scalar variance 
from the flow scale and the nonsingular spectrum is $\sim 1/t^{7/2}$; 
the contribution from the small-$k$ peak is $\sim 1/t^{5/2}$. 
The latter dominates and, therefore, determines the decay law for 
the total variance. 
The behavior in the long-time limit is the 
same as for $C_2>0$. 

\begin{figure}[t]
\centerline{\psfig{file=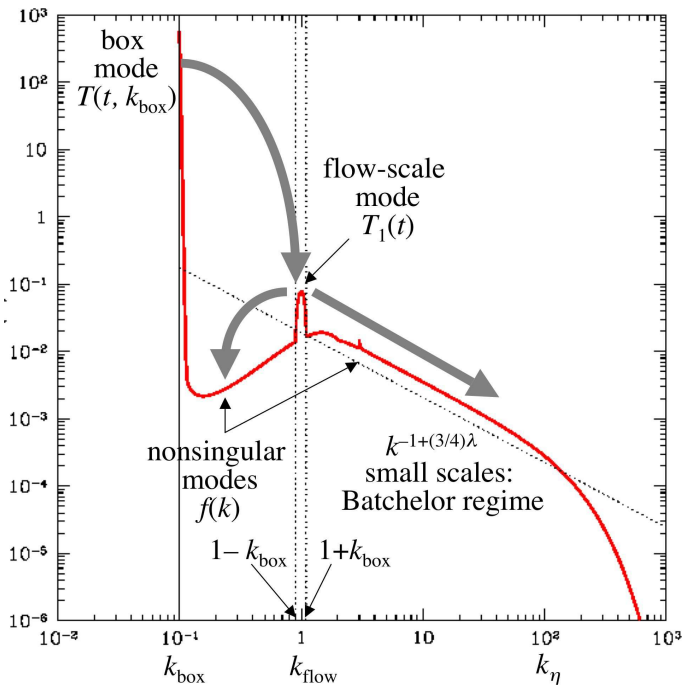,width=9cm}}
\caption{\label{fig_strange_mode} 
Structure of the ``strange mode.'' 
The spectrum is from a run with $\kbox=0.1$, $\keta=100$
(the same as in \figref{fig_Tk}, $t=1000$). 
Gray arrows show directions of couplings 
(transfer of scalar variance).} 
\end{figure}

\subsection{Numerical solution}
\label{sec_num}

We have checked our analytical solution 
by solving \eq{Eq3D} numerically. 
The powerlike decay laws for the flow-scale mode $T(t,1)$ 
and for the total variance are in good agreement 
with theory, but they can only can be seen if the 
scale separation between the box and the flow 
is sufficiently large [$\kbox\sim0.01$; see \figref{fig_Et}(a)]. 
Spectra (\figref{fig_Tk}) 
and long-term decay rates [\figref{fig_Et}(b)] 
are clearly in agreement with theory already 
at moderate scale separations. 
A simple way to estimate the threshold at which the decay rate 
ceases to be determined by box-scale diffusion 
is by requiring that the box decay rate should be smaller 
than the Lagrangian-stretching value: 
$\lambda=(40/9)\kbox^2<1$, so $\kbox\lesssim0.47$. 
Consistent with this estimate, the box value still works 
for $\kbox=1/3$ [\figref{fig_Et}(b); cf.~\cite{Haynes_Vanneste}].
Obviously, \eq{Eq3D} itself with the cutoff 
at $\kbox$ is only technically valid when $\kbox\ll1$, but 
we see that it continues to yield reasonable solutions 
even at moderate $\kbox$. 

\section{Discussion}
\label{sec_disc}

We now have a physical picture of the ``strange mode'': 
the small-$k$ peak (singularity at the box scale) 
serves as a slowly decaying source 
to the flow-scale mode (singularity at $k=1$), 
which, in turn, is mixed by the random flow and 
thus excites the nonsingular modes at small [\eq{FPSln}] 
and large [\eq{T_series}] scales. 
The structure of the spectrum 
is illustrated in \figref{fig_strange_mode}.

The low-wave-number behavior of the decaying 
scalar field was previously analyzed in a heuristic 
way by Kerstein and McMurtry~\cite{Kerstein_McMurtry} 
(see also \cite{Gonzalez} for a treatment based on 
one of the turbulence closure schemes, which gives mostly 
similar results). 
They considered advection by a narrow-band (i.e., single-scale) 
forced random flow in an unbounded domain --- i.e., in 
the regime that we call the transient (powerlike-decay) stage. 
They recognized the defining role of coupling 
between the large scales ($k\lesssim\kpeak$) 
and the flow scale ($k=\ko=1$) and derived the $k^4$ scaling 
at $k\ll\kpeak$ with a exponential fall off at $k>\kpeak$ 
[\eq{HomSln_C2}] and the ensuing powerlike-decay laws 
for the case of $C_2=0$ (see the end of \secref{sec_C2_zero}). 
For the intermediate rage $\kpeak\ll k\ll1$, 
they predicted a $k^2$ spectrum (in 3D) 
--- in contrast to our $k^1$ result 
[\eqsand{Tint_decay_C2}{T_series} and \secref{sec_C2_zero}]. 
The reason for the discrepancy is as follows. 
The analysis of \cite{Kerstein_McMurtry} is based on 
Taylor-expanding the flow around $k=1$ --- i.e., in terms of our 
theory --- setting $T(t,1+q)\simeq T(t,1)$ in \eq{HomEq}, which gives 
$S(t,k)\simeq k^4 T(t,1)$. If we had used the resulting equation 
to solve for $T(t,k)$ at $\kpeak\ll k\ll 1$, we would also 
have obtained $T(t,k)\sim k^2$. However, as we have seen above, 
the width of the flow-scale singularity is $\sim\kpeak$ 
[\eqsand{T1_decay_C2}{T1_decay_zeroC2}], 
so Taylor expansion cannot be used in \eq{HomEq}
for $k\gg\kpeak$. In this intermediate range, 
the integral in \eq{HomEq} must instead 
be replaced by the integral over the entire flow-scale peak, 
resulting in our $k^1$ scaling. The $k^2$ term enters as a 
correction due to the interaction between nonsingular 
modes [\eq{T_series}]. 

Finally, let us comment on our modeling assumptions. 
The white-noise approximation might appear drastic: 
the correlation time of any realistic flow 
is comparable to the flow 
time scale $\sim(u\ko)^{-1}\sim\kappa_2^{-1}$. 
However, since the scalar decay time is much longer 
than the flow time scale (provided $\kbox\ll\ko$), 
the white-noise model appears reasonable. 
We believe it also correctly captures the small-scale 
structure: the key factor here is 
the statistics of fluid displacements, which are 
integrals of velocity and are finite-time correlated 
even for a white-in-time velocity. 

Our model flow was single scale. 
Although such flows can be set up in 
the laboratory \cite{Rothstein_etal,Voth_etal}
\footnote{It was pointed out to us by Kerstein 
\cite{Kerstein_pc} that a single-scale 
random flow could also be set up by randomly stirring 
a granular material: studying mixing in such a flow 
would provide an interesting experimental test.}, 
the real-world mixing problems usually contain 
(at sufficiently small scales) 
a wide (inertial) scale range of three-dimensional 
turbulent motions. 
While the variance spectrum in the inertial range 
should follow the Obukhov-Corrsin 
$k^{-5/3}$ law \cite{Obukhov_scalar,Corrsin_spectrum}
and there will be another transient powerlike-decay stage 
\cite{Corrsin_inv,Chasnov,Eyink_Xin,Chaves_etal,Chertkov_Lebedev}, 
the long-term decay (after the scalar variance reaches $k<\ko$) 
should still be qualitatively described by our theory. 
Another modification that results from the relaxation 
of the single-scale assumption concerns the intermediate 
wave-number range $\kpeak\ll k\ll\ko$. As noted 
in \cite{Kerstein_McMurtry} and confirmed in 
pipe-flow mixing experiments \cite{Guilkey_etal_spectra}, 
the interaction between the $k=\kpeak$ mode and the 
low-wave-number tail of the kinetic-energy spectrum 
($\sim k^4$) can change the scaling of the scalar-variance 
spectrum in this range. 

In conclusion, 
we emphasize that, in any laboratory experiment aiming 
to test our results, the stirring must be done at scales 
substantially smaller than the system size 
to ensure that $\kbox\ll\ko$. It was just such a set up 
(in 2D) that allowed Voth {\em et al.} \cite{Voth_etal} 
to show experimentally that the global mixing rate was much 
smaller than that predicted by the Lagrangian-stretching theories 
and consistent with the box-scale turbulent-diffusion rate 
--- precisely the point the theory presented above 
is meant to demonstrate. 

\begin{acknowledgments}
We thank J.-L.~Thiffeault, A.~Kerstein, and M.~Gonzalez for 
stimulating discussions. 
A.A.S.\ was supported by the Leverhulme Trust through 
the UKAFF. 
\end{acknowledgments}


\bibliography{shc_PRE}

\end{document}